\documentclass[times,twocolumn,final,authoryear]{elsarticle}

\usepackage{jasr}
\usepackage{framed,multirow}

\usepackage{amssymb}
\usepackage{latexsym}

\usepackage[switch]{lineno}

\usepackage{url}
\usepackage{xcolor}
\definecolor{newcolor}{rgb}{.8,.349,.1}
\usepackage{epsfig}
\usepackage{color}
\newcommand{\fdg}{\mbox{\ensuremath{.\!\!^\circ}}}%
\usepackage{tabularx}
\usepackage{multirow}
\usepackage[para,online,flushleft]{threeparttable}
\usepackage[ampersand]{easylist}
\usepackage{graphicx}	
\usepackage{amsmath}	
\usepackage{amssymb}	
\usepackage[switch]{lineno}
\linenumbers
\usepackage[citebordercolor=white]{hyperref}

\usepackage{hyperref}
\hypersetup{
    colorlinks=true,
    linkcolor=blue,
    filecolor=blue,      
    urlcolor=blue,
    citecolor=blue,
    filecolor=blue,
}
\journal{Advances in Space Research}

\begin{document}

\verso{E.~Aktekin}

\begin{frontmatter}

\title{Investigation of X-ray emission from the unidentified TeV gamma-ray source HESS~J1832$-$085 with {\it Suzaku}}%
\tnotetext[tnote1]{Corresponding author.}

\author[1]{Ebru Aktekin\corref{cor1}}
\cortext[cor1]{E-mail: ebrucaliskan@sdu.edu.tr}

\affiliation[1]{organization={Department of Physics},
                addressline={S\"{u}leyman Demirel University},
                city={Isparta},
                postcode={32000},
                country={Turkey}}

\received{xx}
\finalform{xx}
\accepted{xx}
\availableonline{xx}
\communicated{}

\begin{abstract}
Observations conducted with H.E.S.S. at high energies have led to the discovery of numerous gamma-ray sources in the Galactic plane at TeV energies. One of these sources, HESS~J1832$-$085, has been suggested to be a pulsar wind nebula (PWN); however, its nature is not yet fully understood. In this work, we analyze {\it Suzaku} data to investigate the X-ray spectral properties of HESS~J1832$-$085. We found that the X-ray spectra are highly absorbed and well-represented by a power-law model with a photon index of $\Gamma \sim 1.5$, and an unabsorbed X-ray flux of $F_{\rm X}  \sim 0.3 \times 10^{-11}$~erg~cm$^{-2}$~s$^{-1}$ in the 2$-$10 keV energy band. The gamma-ray flux is approximately 66 times higher than the X-ray flux. Based on our X-ray analysis, we discuss the origin of the source HESS~J1832$-$085. We propose that the PWN scenario is possible, although several issues still need to be resolved.

\end{abstract}


\end{frontmatter}

\nolinenumbers

\section{Introduction}
In recent years, observations with the High Energy Stereoscopic System (H.E.S.S.), a system designed to detect very high energy (VHE; 0.1$-$100 TeV) gamma rays, have led to the discovery of numerous gamma-ray sources at TeV energies along the Galactic plane. Some newly discovered TeV objects lack counterparts at other wavelengths and are referred to as unidentified TeV gamma-ray sources. These objects play a crucial role in studying the origins of high-energy cosmic rays. Multiwavelength observations are essential to understand their emission mechanisms and achieve accurate identification. Especially the observations in X-rays offer deeper insights into which particles are being accelerated and their associated dynamics. In typical interstellar magnetic fields with strengths of a few $\mu$ Gauss, high-energy electrons primarily produce synchrotron X-rays. Consequently, the flux ratio between X-rays and TeV gamma-rays serves as a key indicator to determine if the accelerated particles are protons or electrons (e.g. \citealt{Matsumoto2007, Bamba2007}). 

The source HESS~J1832$-$085 was discovered through observations conducted at TeV energies by H.E.S.S. \citep{He18}. It exhibits a point-like morphology, with an extension measured to be less than 0.05 degrees. Its photon flux is almost 0.8 per cent of that of the Crab Nebula, and it has a photon index of $\Gamma = 2.38 \pm 0.14$ \citep{He18}. The pulsar PSR J1832$-$0827 \citep{Cl86} is spatially coincident with HESS~J1832$-$085 \citep{He18}. The distance to PSR~J1832$-$0827 was estimated to be $\sim$4.9~kpc \citep{Co02} and $4.4-6.1$~kpc \citep{Fr91}. 

\citet{Ma19} investigated the SNR candidate G23.11+0.18 using the Murchison Widefield Array (MWA) radio continuum data. They proposed that its distance is $4.6 \pm 0.8$~kpc and that its progenitor is a wind-blown bubble. The authors also discussed a potential connection between HESS~J1832$-$085 and suggested that it is likely associated with G23.11+0.18. Using {\it Fermi}-LAT data, \citet{Er21} reported the detection of excess GeV gamma-ray emission that partially overlaps with both G23.11+0.18 and HESS~J1832$-$085 in the northern region of the SNR.

Nevertheless, no definitive conclusion was reached regarding the pulsar wind nebula (PWN) scenario due to the absence of a known PWN counterpart at other wavelengths. The nature of HESS~J1832$-$085 remains unclear. In this work, we investigate the nature of the emission from the source HESS~J1832$-$085 using {\it Suzaku} \citep{Mi07} data with the high spectral resolution X-ray Imaging Spectrometer (XIS: \citealt{Ko07}). The details of the observation and the data reduction method are presented in Section \ref{Sec2}, and the findings obtained from the analysis are presented in Section \ref{Sec3}. The results are discussed in Section \ref{Sec4}.

\section{Observation and data reduction}
\label{Sec2}
We retrieved the {\it Suzaku} X-ray data of HESS~J1832$-$085 from the \texttt {HEASARC} public databases\footnote{\url{https://heasarc.gsfc.nasa.gov/db-perl/W3Browse/w3browse.pl}}. We analyzed data taken with the XIS, which include XIS0, XIS1, and XIS3 CCDs. General information about the XIS observation is presented in Table~\ref{Table1}.

The data were analyzed with the \texttt {HEASoft}\footnote{\url{https://heasarc.gsfc.nasa.gov/docs/software/heasoft/}} package v.6.29 and \texttt {xspec} \citep{Ar96}. The $\chi^{2}$ statistic was used in the spectral analysis.

\begin{table*}
\label{Table1}
\begin{minipage}{170mm}
\begin{center}
 \caption{Observation log.} 
\begin{tabular}{@{}p{3.0cm}p{2.0cm}p{1.8cm}p{2.8cm}p{2.2cm}p{2.2cm}@{}}
 \hline
Observation name       			& Obs. ID 	          & Start time & Pointing direction        	&  Exposure time   & Obs. PI         \\ 
            			&                     &                 & ($l, b$) &     (ks)           \\
\hline
HESSJ1832 		        &     506021010          & 2011-04-08         &  ($23\fdg299$, $0\fdg310$) & 40.3   &	H. Matsumoto          \\ 
\hline
\label{Table1}
\end{tabular}
\end{center}
\end{minipage}
\end{table*}

\section{Analysis and results}
\label{Sec3}
\subsection{X-ray image}
\begin{figure*}
\centering
\includegraphics[width=12cm]{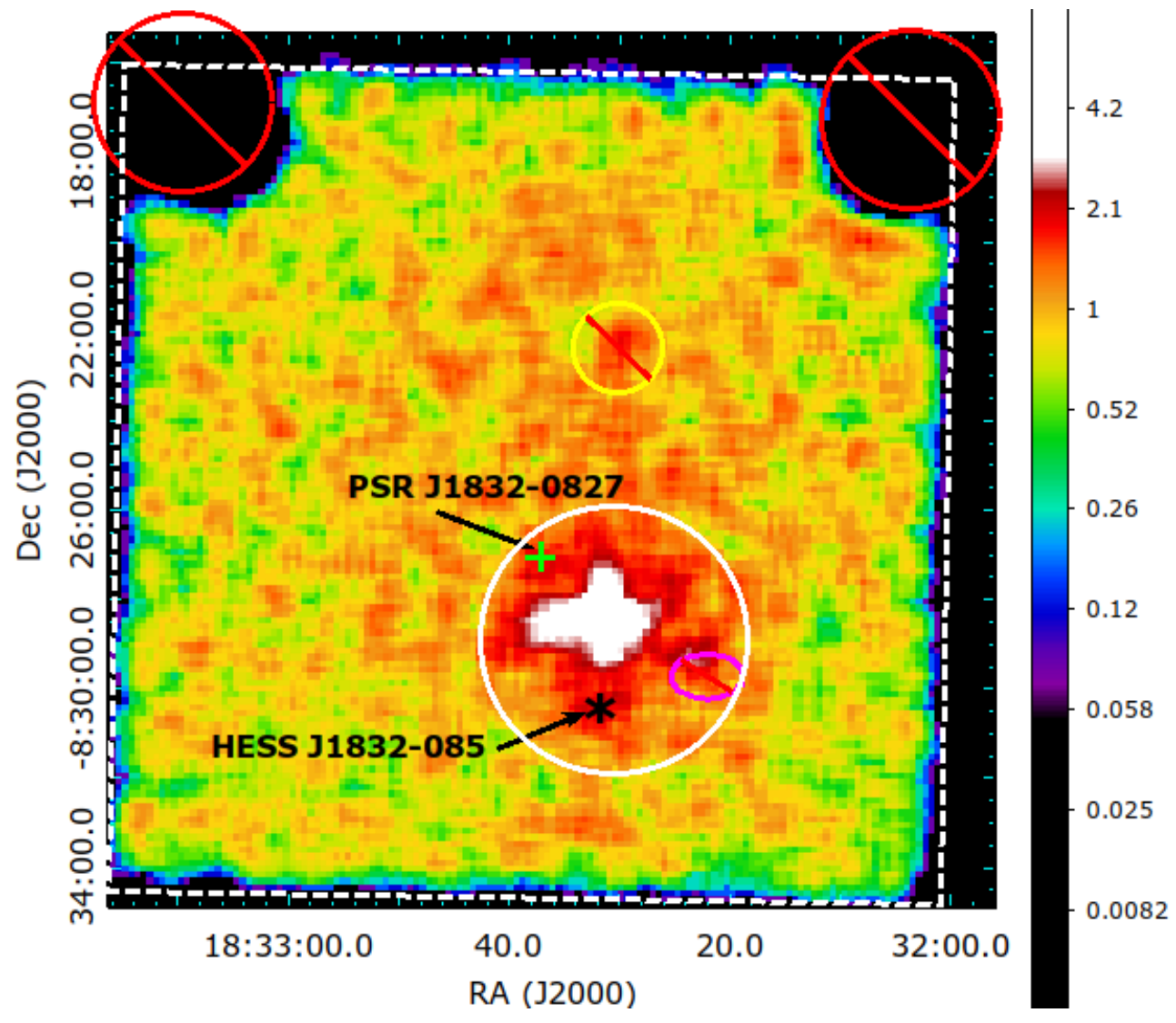}
\includegraphics[width=8cm]{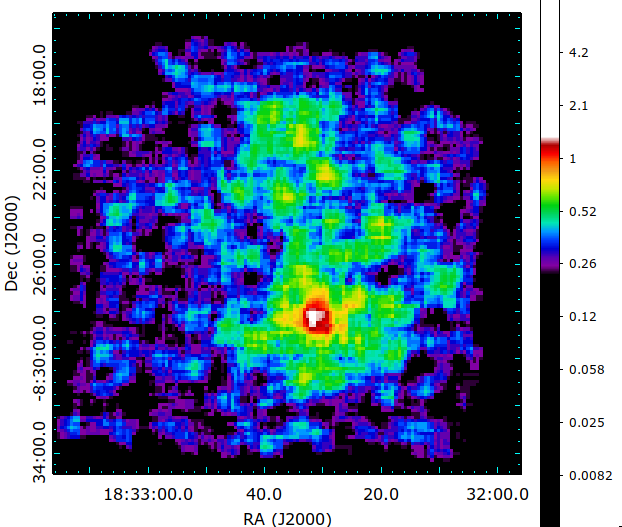}
\includegraphics[width=8cm]{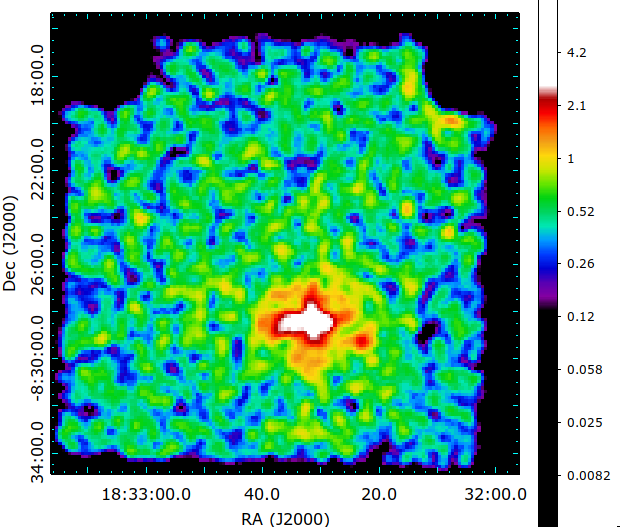}
\caption{\textbf{Upper Panel:} {\it Suzaku} XIS1 image of HESS~J1832$-$085 in the 0.3$-$10.0 keV energy range. The marks $\ast$ and \texttt {+} indicate the center locations of HESS~J1832$-$085 \citep{He18} and PSR J1832$-$0827 \citep{Cl86}, respectively. The area indicated by a white circle represents the source emission, while the dashed square shows the area selected for background analysis. Two sources, 2MASS~J18322422$-$0829300 (magenta ellipse) and SPICY 85740 (yellow circle), are also excluded from the analysis. \textbf{Lower panels:} Soft (1.0$-$3.0 keV) (left) and hard (3.0$-$10.0 keV) (right) X-ray images of HESS~J1832$-$085 obtained with {\it Suzaku} XIS1. NXB subtraction and vignetting correction are performed for all images. The calibration sources in the corners are excluded from all images.}
\label{figure1}
\end{figure*}

The X-ray image of the source HESS~J1832$-$085 in the 0.3$-$10.0 keV energy range has been created and is presented in Fig.~\ref{figure1}. In the lower left and right panels of Fig.~\ref{figure1}, we also showed the soft (1.0$-$3.0 keV) and hard (3.0$-$10.0 keV) band images of HESS~J1832$-$085, respectively. In the 0.3$-$10 keV energy range X-ray image, the source HESS~J1832$-$085 exhibits extended emission within an area of approximately 3~arcmin in radius and a peak position of $l$=23\fdg237, $b$=0\fdg313. It appears bright in hard X-ray image (see Fig.~\ref{figure1}, lower right panel). 

\begin{figure}
\centering
\includegraphics[width=9cm]{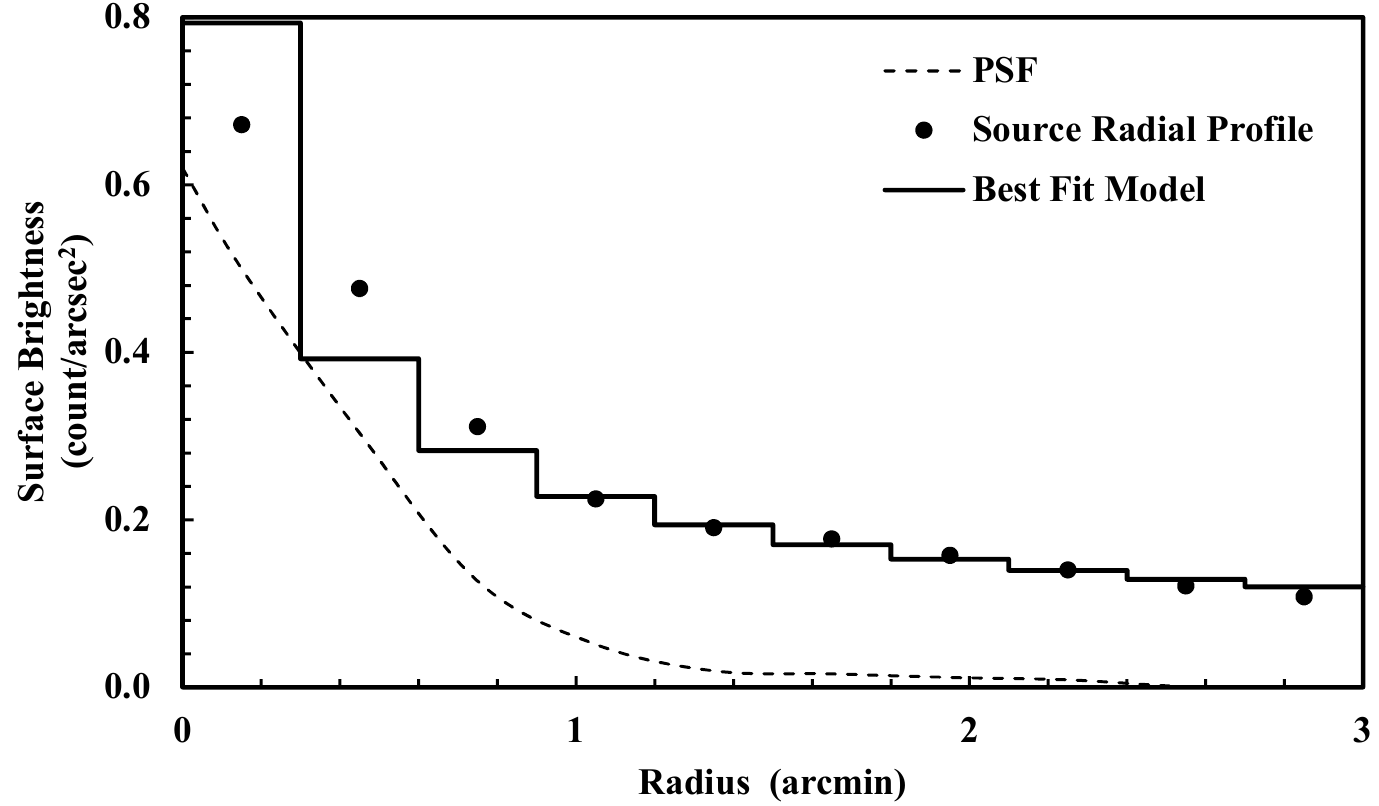}
\caption{Radial profile of HESS~J1832$-$085 and the PSF in the energy range of 0.3$-$10.0 keV.  The source's radial profile is shown by circles, the solid line indicates the best-fit  ($\chi^{2}$/d.o.f.=0.997), and the dashed line displays the instrument's PSF.}
\label{figurepsf}
\end{figure}

Although {\it Suzaku} has a low and stable background, the point spread function (PSF) of the {\it Suzaku} X-ray telescope is approximately 2 arcmin \citep{Serlemitsos2007}, which is broader than those of {\it Chandra} and {\it XMM-Newton}. Consequently, we compared the observed radial profile with the PSF simulated using \texttt {xissim}. We fitted the 0.3$-$10.0 keV radial profile using the PSF plus a constant representing the background. The fit was rejected with a $\chi^{2}$/d.o.f. of 26.5/6. We found large residuals extending beyond the PSF and concluded that the emission is indeed extended. The radial profile of the source and the instrument's PSF is displayed in Fig. ~\ref{figurepsf}.

According to \citet{Ma19}, the MWA radio continuum image indicates a shell-like morphology for the SNR candidate G23.11+0.18 ($l$=23\fdg12, $b$=0\fdg19, radius\footnote{\citet{Ma19} estimated the radius of G23.11+0.18 to be 700$\pm$50 arcsec.} of about 750~arcsec), which was initially detected in the radio band by \citet{Anderson2017}. However, \citet{Ma19} also reported that their analysis of the {\it XMM-Newton} image showed no significant diffuse X-ray emission, making it impossible to provide additional evidence supporting G23.11+0.18 using this X-ray dataset. They attributed this to a likely high level of foreground absorption.

\subsection{Spectral analysis}
For spectral analysis, a circular region with a radius of 3~arcmin was selected, as shown in Fig.~\ref{figure1}. For background analysis, an area covering the entire field of view (FoV) of the XIS was chosen, and the source emission was subtracted from this area. Because the source is located along the Galactic plane, the local Galactic emission and Cosmic X-ray Background (CXB) must be taken into account. For background estimation, we considered the non-X-ray background (NXB), CXB, Galactic Ridge X-ray Emission (GRXE), and Local Hot Bubble (LHB). The background spectra were subtracted from the source spectra.

Since the X-ray emission from a PWN can be well described by a power-law model, we first applied a power-law model to the spectra. We modeled the absorption with the \texttt {TBabs} model \citep{Wi00}. 

In this fitting (\texttt {TBabs*power-law}), the hydrogen column density ($N_{\rm H}$), photon index ($\Gamma$), and normalization were treated as free parameters. We obtained $N_{\rm H} = 5.88^{+1.12}_{-0.99}\times10^{22}$ cm$^{-2}$ at a 90 per cent confidence level  with $\chi^{2}_{\nu}$ =1.04 (d.o.f.=202). $N_{\rm H}$  value falls within the expected range for sources located in the Galactic plane. We also calculated the total $N_{\rm H}$ using \texttt {nhtot}\footnote{\url{https://www.swift.ac.uk/analysis/nhtot/index.php}} which includes contributions from both atomic and molecular \texttt {H} \citep{Willingale2013}. We found $N_{\rm H}$  $\sim 1.7\times10^{22}$ cm$^{-2}$, which is nearly three times less than the value obtained from the spectral fit. This discrepancy persists even when accounting for the error range of the $N_{\rm H}$ value obtained from the spectral fit. The high absorption supports the idea that the source is embedded in dense gas, as molecular clouds in this direction were reported by \citet{Ma19}. Some sources, such as HESS~J1804$-$216 \citep{Bamba2007} and HESS~J1813$-$178 \citep{Brogan2005}, also show evidence of excess absorption.

\begin{figure*}
\centering
\includegraphics[width=14cm]{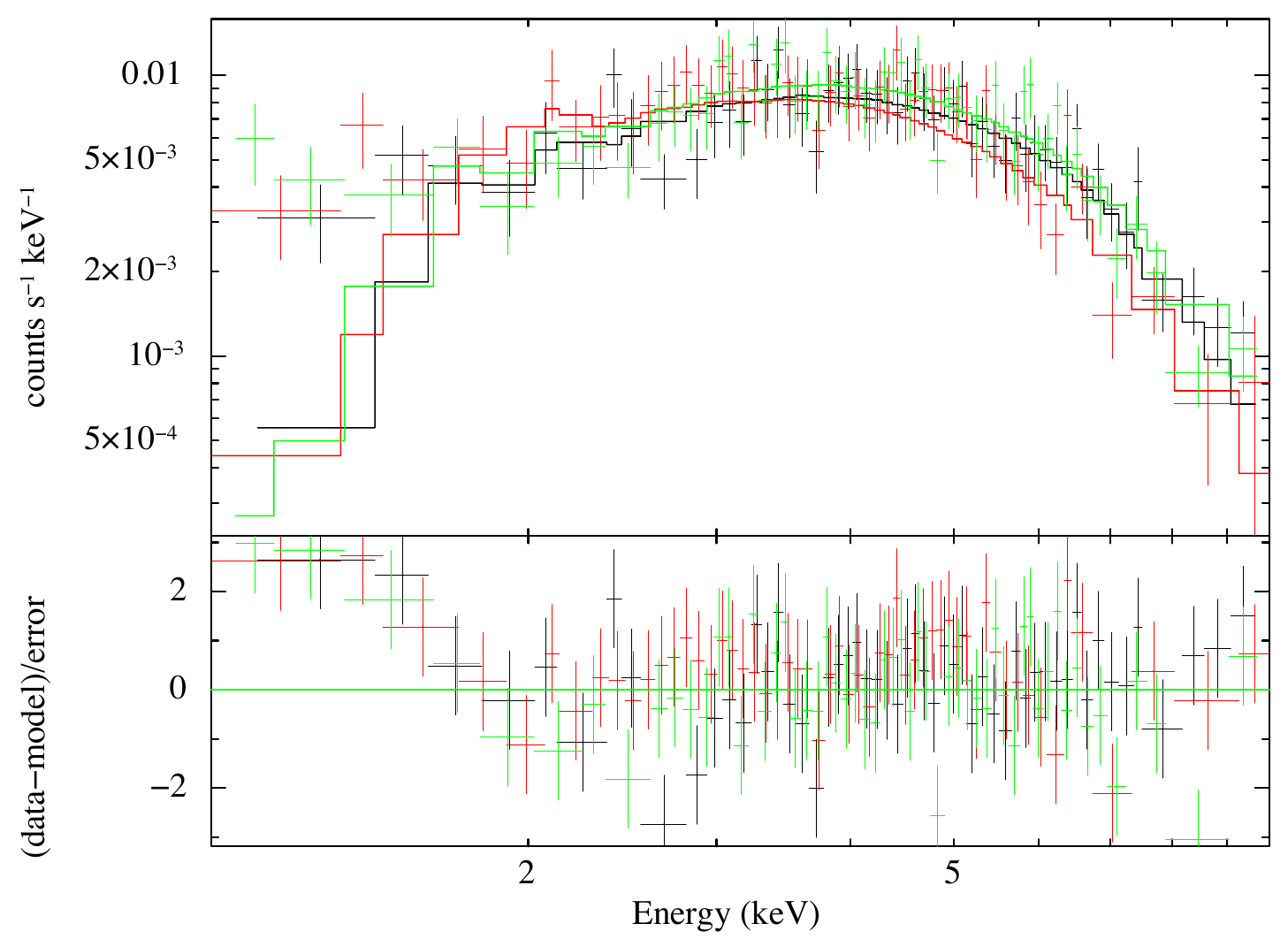}
\caption{{\it Suzaku} XIS spectra (XIS0: black, XIS1: red, and XIS3: green) of HESS~J1832$-$085 in the 1.0$-$10.0 keV energy band. The best-fit model (an absorbed \texttt{power-law}) is plotted as solid lines.}
\label{figure2}
\end{figure*}

The {\it Suzaku} XIS spectra in the 1.0$-$10.0 keV energy band are presented in Fig.~\ref{figure2}. The model curves appear wavy in the 3$-$6 keV band, indicating an excess of high-energy photons, which can be better fit by adding a second, higher temperature component. To investigate whether there is a thermal component in the spectra, the \texttt {vpshock} model was added to the \texttt {power-law} model. The free parameters included the $N_{\rm H}$, electron temperature ($kT_{\rm e}$), upper limit of the ionization time-scale ($\tau_{\rm u}$), and normalization. The model yielded a very low temperature ($kT_{\rm e}$ $\sim$ 0.1 keV) and an unrealistic error of the upper limit on the ionization timescale, with $\chi^{2}_{\nu}$ = 0.89 (d.o.f.= 199). Allowing the metal abundances to vary as free parameters did not improve the fit. We also attempted to fit the spectra using a pure thin-thermal plasma model, such as \texttt{nei} model. In this case, the problem with residuals below 2 keV was resolved, and a suitable reduced $\chi^{2}$ value was achieved. However, this approach yielded a very low temperature ($kT_{\rm e}$ $\sim$ 0.1 keV) and unrealistic errors of parameters. Thus, we conclude that a thermal interpretation of the spectra is unlikely and instead favor a non-thermal explanation.

The best-fit parameters obtained from the {\it Suzaku} X-ray spectral analysis  are presented in Table~\ref{Table2}.  We also estimated the absorbed and unabsorbed fluxes, along with their uncertainties, for the  X-ray spectra in the 2$-$10 keV energy range using the \texttt {xspec} model \texttt {cflux} (see Table~\ref{Table2}).

\begin{table*}
\centering
\caption{Fitting parameters of the XIS spectra. The error ranges for each parameter are shown in parentheses at a 90 per cent confidence level.}
\begin{tabular}{@{}p{4.2cm}p{4.8cm}p{4.8cm}@{}}
\hline
Model  & Parameter (Unit)  & Value   \\
\hline \\[-2.0ex]
TBabs     &  $N_{\rm H}$ ($10^{22}$ cm$^{-2})$          &  5.88 (4.89$-$7.00)   \\ [0.25 cm] 
Power-law  &      Photon index, $\Gamma$                &  1.54 (1.35$-$1.74) \\ [0.25 cm] 
           &      Norm$^{\dagger}$  ($10^{-3}$)         &  0.55  (0.39$-$0.78) \\ [0.25 cm] 
           &      Absorbed flux$^{\ddagger}$             &  0.22 (0.21$-$0.23) \\ [0.25 cm] 
           &      Unabsorbed flux$^{\ddagger}$           &  0.29 (0.28$-$0.30)  \\ [0.25 cm] 

& $\chi^{2}_{\nu}$ (d.o.f.)                & 1.04 (202) \\
\hline
\end{tabular}
\label{Table2}
\begin{flushleft}
\item $^{\dagger}$ The norm is in units of photons~cm$^{-2}$~s$^{-1}$~keV$^{-1}$ at 1~keV. \\
\item $^{\ddagger}$ The flux is in $10^{-11}$~erg~cm$^{-2}$~s$^{-1}$ in the 2$-$10 keV energy band. \\
\end{flushleft}
\end{table*}

\subsection{Light curve}
\begin{figure*}
\centering
\includegraphics[width=9.1cm]{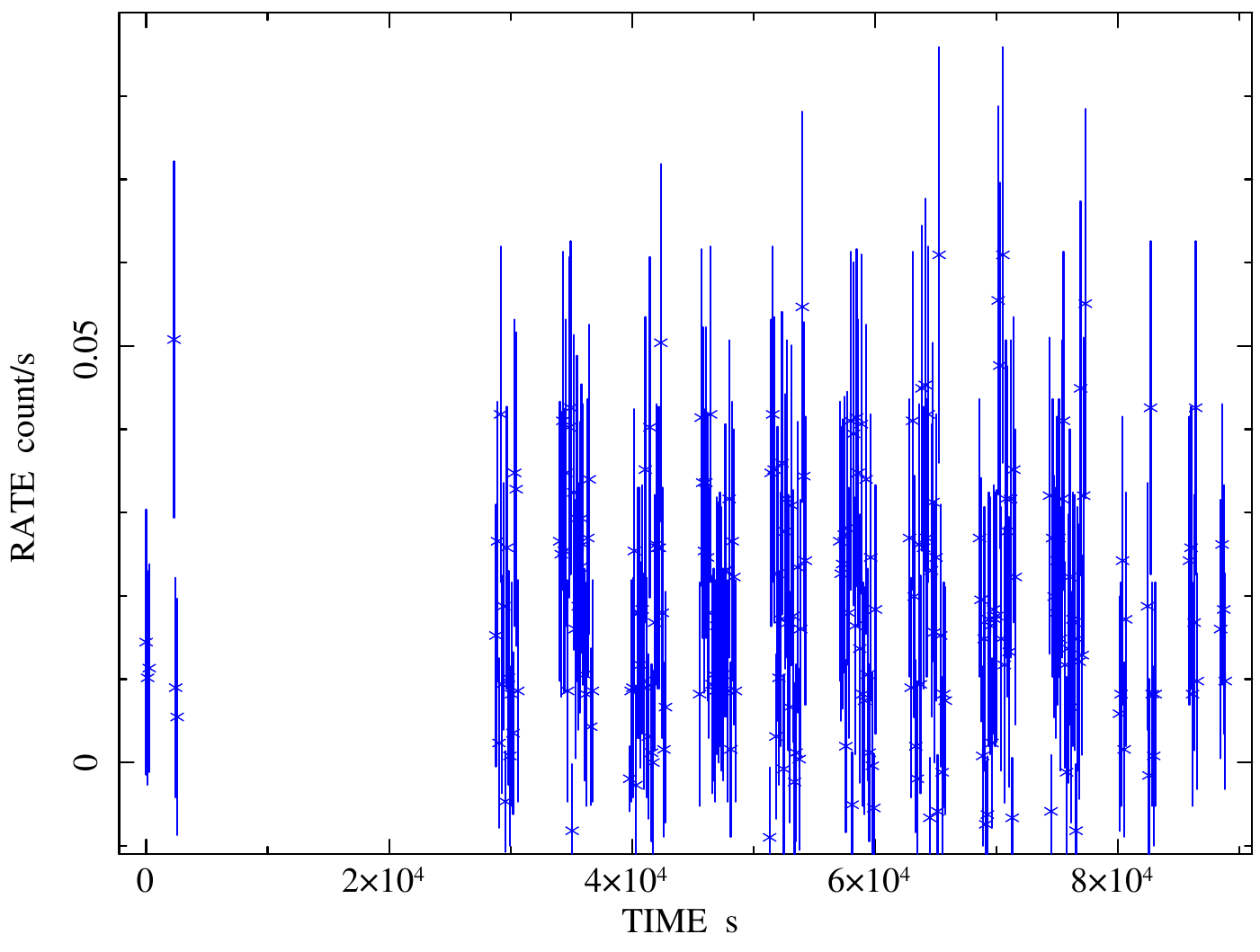}
\includegraphics[width=9.1cm]{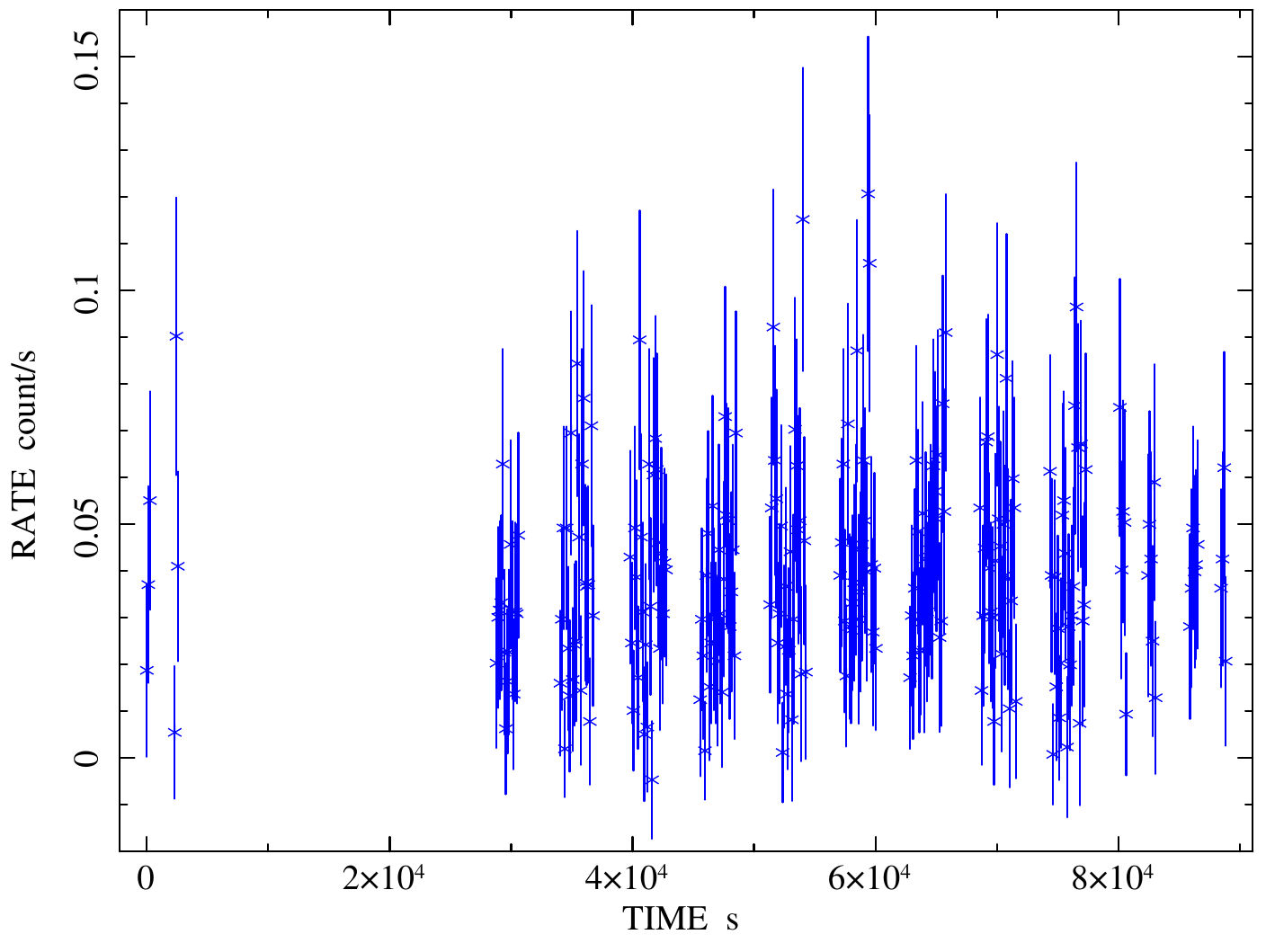}
\caption{The background-subtracted light curves of HESS~J1832$-$085 in the 1.0$-$3.0 keV energy band (left) and in the 3.0$-$10.0 keV energy band (right), obtained with {\it Suzaku} XIS1. The time bins are 128s. The light curve in the soft band exhibits a low count rate.}
\label{figure3}
\end{figure*}

A variability is more commonly seen in other types of compact, X-ray-emitting Galactic sources, such as X-ray binaries or cataclysmic variable stars. We obtained light curves for the source HESS~J1832$-$085 in both the soft (1.0$-$3.0 keV) and hard (3.0$-$10.0 keV) energy bands, as shown in Fig.~\ref{figure3}.  
    
Using \texttt {lcstats}\footnote{\url{https://heasarc.gsfc.nasa.gov/ftools/fhelp/lcstats.html}} (v.1.0) tool of \texttt {XRONOS} (v.6.0), we found the root mean square fractional variation for the 1.0$-$3.0 keV light curve to be less than 0.293 at the 3-$\sigma$ level, and for the 3.0$-$10.0 keV light curve to be $0.238\pm0.042$. This result indicates that the {\it Suzaku} light curve does not show significant time variability in either band. \\

\section{Discussion}
\label{Sec4}
In this work, the X-ray nature of the source HESS~J1832$-$085 was investigated with {\it Suzaku} data. Using the obtained X-ray properties, the possibility of HESS~J1832$-$085 being a PWN was investigated. The results obtained are discussed below.

We investigated the extended source scenario, considering the correlation between the size of the extension and the characteristic age of the associated pulsar (e.g. \citealt{Bamba2010}). \citet{Bamba2010} studied the relationship between the characteristic ages of host pulsars and the X-ray sizes of PWNe, demonstrating that the extent of X-ray emission increases with the pulsar's characteristic age. They reported that the size of the extended emission from PWNe continues to expand for up to approximately 100 kyr as the characteristic age increases. The angular X-ray size of  HESS~J1832$-$085 (3 arcmin) corresponds to approximately 4.3 pc ($d$/4.9 kpc). The characteristic age of PSR J1832$-$0827 (161 kyr; see Table 1 of \citealt{Kargaltsev2013}) is significantly larger than those of other sources in their sample with similar X-ray sizes.

The pulsar PSR J1832$-$0827 is located near ($\sim$1.96 arcmin in projection; see Fig.~\ref{figure1}) the northeastern part of HESS~J1832$-$085, making it possible that PSR J1832$-$0827 is related to it. However, the uncertainty in the distance to HESS~J1832$-$085 prevents a definitive conclusion about this association.

We found the X-ray spectra of HESS~J1832$-$085 are well described by a \texttt {power-law} model with a photon index of $\Gamma \sim 1.5$. No line emission from any elements was detected in the spectra. Therefore, we concluded that the X-ray emission from the source is synchrotron emission. 

The photon index value ($\Gamma \sim 1.5$) we obtained is within the expected range for PWN ($\Gamma \sim 2$: \citealt{Ga06}; $\Gamma \sim 1-2$: \citealt{Ka08}). This result suggests that HESS~J1832$-$085 could be a PWN. 

We compare unabsorbed X-ray flux $F_{\rm X} = (0.29 \pm 0.01) \times 10^{-11}$ erg~cm$^{-2}$~s$^{-1}$  with gamma-ray flux $F_{\rm TeV} = 0.8$ per cent of the Crab \citep{He18} using the formula given by \citet{Ya06}

\begin{equation}
R_{\rm TeV/X}=\frac{F_\gamma(1-10~{\rm TeV})}{F_X(2-10~{\rm keV})}~~,
\end{equation}

and obtained $R_{\rm TeV/X}$ = ($19.2 \pm 0.2$)$\times$ 10$^{-11}$/($0.29 \pm 0.01$) $\times$ 10$^{-11}$ $=$ $66.2 \pm 1.6$. This high flux ratio is consistent with a hadronic scenario but is unlikely for PWNe \citep{Ya06}.\\

The X-ray luminosity ($L_{\rm X}$) of PWNe is known to correlate with the spin-down energy ($\dot{E}$) of their associated pulsars (e.g. \citealt{Ka08}). For HESS~J1832$-$085, we estimated an unabsorbed X-ray luminosity of $L_{\rm X} \sim 0.83 \times 10^{34}$~erg~s$^{-1}$ in the 2$-$10~keV energy band, assuming a distance of 4.9~kpc, with a 90\% confidence range. Using this value, we calculated the X-ray efficiency ($\eta$), defined as $L_{\rm X} / \dot{E}$, to be approximately 0.83, based on the spin-down energy of the PSR J1832$-$0827, $1 \times 10^{34}$~erg~s$^{-1}$ \citep{Kargaltsev2013, Ma19}. This efficiency is higher than the typical range observed for neutron stars, where $\eta$ typically lies between $10^{-5}$ and $10^{-1}$ (see Figure 5 of \citealt{Ka08}). 

From the {\it Suzaku} light curve (see Fig.~\ref{figure3}), we did not observe significant time variability in either the soft or hard bands. This lack of variability supports the PWN scenario. \\

In conclusion, we have presented the first {\it Suzaku} analysis of HESS~J1832$-$085. To understand its nature, we examined its X-ray images, spectra, and light curves. We propose that the PWN scenario as the origin of HESS~J1832$-$085 is possible, although there are several issues to be solved. Multi-wavelength analyses of this source, particularly X-ray observations with high-energy resolution satellites such as {\it XRISM} \citep{Tashiro2020} and {\it Athena} \citep{Barret2018} are needed to confirm its PWN nature.

\section*{Acknowledgments}
The author thanks the reviewers for their constructive comments and suggestions that have improved this paper. The author also thanks all the members of the {\it Suzaku} team for their support in the observation, software development, and calibration processes. 

\newpage
\bibliographystyle{jasr-model5-names}
\biboptions{authoryear}
\bibliography{refs}

\end{document}